\documentclass{ieeeaccess}

\usepackage{cite}
\usepackage{amsmath,amssymb,amsfonts}
\usepackage{algorithmic}
\usepackage{graphicx}
\DeclareGraphicsExtensions{.pdf,.jpeg,.png,.jpg}
\usepackage{textcomp}
\usepackage{epstopdf}
\usepackage{bm}
\usepackage{subfig}
\usepackage{url}
\usepackage{upgreek}
\usepackage{mathrsfs}

\def\BibTeX{{\rm B\kern-.05em{\sc i\kern-.025em b}\kern-.08em
    T\kern-.1667em\lower.7ex\hbox{E}\kern-.125emX}}

\begin{document}

\history{Received June 4, 2018, accepted August 21, 2018, date of publication September 3, 2018, \textcolor{red}{date of current version November, 2024.}}
\doi{0.1109/ACCESS.2018.2868268}

\title{Micro Hand Gesture Recognition System Using Ultrasonic Active Sensing}

\author{\uppercase{Yu Sang}\authorrefmark{1}, 
\uppercase{Laixi Shi\authorrefmark{2}, 
	and Yimin Liu}\authorrefmark{3}
\IEEEmembership{Member, IEEE}}

\address[1]{Department
		of Electronic Engineering, Tsinghua University,
		Beijing, 100086 China }
\address[2]{Department
	of Electronic Engineering, Tsinghua University,
	Beijing, 100086 China }
\address[3]{Department
	of Electronic Engineering, Tsinghua University,
	Beijing, 100086 China }

\tfootnote{\textcolor{red}{Revised November 28, 2024; This is a updated arxiv version of the publiced IEEE Access version.}}

\markboth
{Yu Sang \headeretal: Micro Hand Gesture Recognition System Using Ultrasonic Active Sensing}
{Yu Sang \headeretal: Micro Hand Gesture Recognition System Using Ultrasonic Active Sensing}

\corresp{Corresponding author: Yimin Liu (e-mail: yiminliu@tsinghua.edu.cn).}

\begin{abstract}
	In this paper, we propose a micro hand gesture recognition system and methods using ultrasonic active sensing. This system uses micro dynamic hand gestures for recognition to achieve human-computer interaction (HCI). The implemented system, called hand-ultrasonic-gesture (HUG), consists of ultrasonic active sensing, pulsed radar signal processing, and time-sequence pattern recognition by machine learning. We adopt lower-frequency (300 kHz) ultrasonic active sensing to obtain high resolution range-Doppler image features. Using high quality sequential range-Doppler features, we propose a state-transition-based hidden Markov model for gesture classification. This method achieves a recognition accuracy of nearly 90\% by using symbolized range-Doppler features and significantly reduces the computational complexity and power consumption. Furthermore, to achieve a higher classification accuracy, we utilize an end-to-end neural network model and obtain a recognition accuracy of 96.32\%. In addition to offline analysis, a real-time prototype is released to verify our method's potential for application in the real world.
\end{abstract}


\titlepgskip=-15pt

\maketitle

\section{Introduction}
\PARstart{R}{ecently}, human-computer interaction (HCI) has become increasingly more influential in both research and daily life \cite{kjeldskov2003review}. In the area of HCI, human hand gesture recognition (HGR) can significantly facilitate many applications in different fields, such as driving assistance for automobiles \cite{pickering2007research}, smart houses \cite{pu2013whole}, wearable devices \cite{gong2017pyro}, virtual reality \cite{fingo.org}, etc. In application scenarios where haptic controls and touch screens may be physically or mentally limited, such as during driving, hand gestures, as a touchless input modality, are more convenient and attractive to a large extent.

Among hand gestures, micro hand gestures, as a subtle and natural input modality, have great potential for many applications, such as wearable, portable, mobile devices \cite{ye2014current}\cite{wang2016interacting}. In contrast with wide-range and large shift actions, such as waving or rolling hands, these micro hand gestures only involve the movements of multiple fingers, such as tapping, clicking, rotating, pressing, and rubbing. They are exactly the ways in which people normally use their hands in the real physical world. Directly using the hand and multiple fingers to control devices is a natural, flexible, and user-friendly input modality that requires less learning effort.

In the hand gesture and micro hand gesture recognition area, various techniques have been applied. Numerous optical methods, which are related to computer vision, utilize RGB cameras \cite{chen2003hand}\cite{yao2015hand}, and depth cameras, such as infrared cameras \cite{ren2011depth}\cite{suarez2012hand}. These camera-based methods have been proposed, and some, such as the Microsoft Kinect or Leap Motion, have even been commercialized. These optical approaches that use RGB images or depth information also perform well in subtle finger motion tracking \cite{fingo.org}\cite{song2014air}. However, to our best knowledge, their reliability in harsh environments, such as in the nighttime darkness or under direct sunlight, is an issue owing to the variance of light conditions. It is also not energy efficient for these approaches to achieve continuous real-time hand gesture recognition for HCI, as required for wearable and portable devices \cite{fingo.org}. In addition to camera-based sensing, there are other passive sensing methods such as pyroelectric infrared sensing and WiFi signal sensing. These methods are utilized in HGR and have achieved great effects \cite{gong2017pyro}\cite{zhang2016mudra}.

In comparison, active sensing is more robust under different circumstances because probing waveforms can be designed. Technologies of sound \cite{gupta2012soundwave}, magnetic field \cite{chen2013utrack} and radio frequency (RF) \cite{wang2016interacting} active sensing have begun to attract interest. Magnetic sensing achieved a high precision in finger tracking but required fingers to be equipped with sensors \cite{chen2013utrack}. Radar technologies based on sound or radio frequency (RF) active sensing can obtain range profile, Doppler and angle of arrival (AOA) features from the received signal. Such features are competitive and suitable for use in identifying the signature of hand gestures \cite{wang2016interacting}\cite{li2017sparsity}\cite{molchanov2015short}. The range resolution is determined by the bandwidths of RF waves \cite{skolnik1962introduction}. However, generating RF waves with large enough bandwidths to meet the resolution demand for micro hand gestures recognition costs a lot. Compared to RF radar, active sensing using sound has advantages that can meet the high range and velocity resolution demand owing to the low propagation velocity \cite{wang2016device}\cite{nandakumar2016fingerio}\cite{timmurphy.org}. Furthermore, the hardware of active sensing using sound can be integrated and miniaturized with MMIC or MEMS techniques with low system complexity, thus having great potential to be used in wearable and portable devices \cite{timmurphy.org}. As introduced in \cite{przybyla20153d}, ultrasonic active sensors with MEMS technique usually consume less energy than CMOS image sensors.

In this work, we build a system consisting of active ultrasonic transceiver, range-Doppler feature extraction processing and time-sequence pattern recognition methods to recognize different kinds of micro hand gestures. We named this system HUG (hand-ultrasonic-gesture). The HUG system transmits ultrasonic waves and receives echoes reflected from the palms and fingers. After that, we implement a pulsed Doppler radar signal processing technique to obtain time-sequential range-Doppler features from ultrasonic waves, measuring objects' precise distances and velocities simultaneously through a single transceiving channel. Making use of the high-resolution features, we propose a state-transition based hidden Markov model (HMM) approach for micro dynamic gesture recognition and achieve a competitive accuracy. For comparison and evaluation, we also implement methods such as random forest, a convolutional neural network (CNN), a long short-term memory (LSTM) neural network, and an end-to-end neural network. A brief introduction of this work was given in \cite{godrich2017students}. In this paper, we give the details and further performance evaluation of the HUG system. The main contributions of this work are as follows:
\begin{itemize}
	\item A micro hand gesture recognition system using ultrasonic active sensing has been implemented. Considering the high-resolution range-Doppler features, we propose a state transition mechanism-based HMM to represent the features, significantly compressing the features and extracting the most intrinsic signatures. This approach achieves a classification accuracy of 89.38\% on a dataset of seven micro hand gestures. It also greatly improves the computational and energy efficiency of the recognition part and provides more potential for portable and wearable devices.
	\item With higher resolution of the range-Doppler features, the end-to-end neural network model has been implemented, achieving a classification accuracy of 96.34\% on the same dataset. Furthermore, we have released a real-time micro hand gesture  recognition prototype to demonstrate the real-time control of a music player to verify our system's real-time feasibility.
\end{itemize}

The rest of the paper is organized as follows: in Section II, related works performed in recent years are discussed. In Section III, we formally describe our micro hand gesture recognition system HUG, including the system design, signal processing technologies and machine learning methods for recognition. In Section IV, we introduce the experiments designed for our system, including implementation and performance evaluation. We finally conclude our work, followed by future work and acknowledgments.

\begin{figure}[t]
	\includegraphics[width=0.5\textwidth]{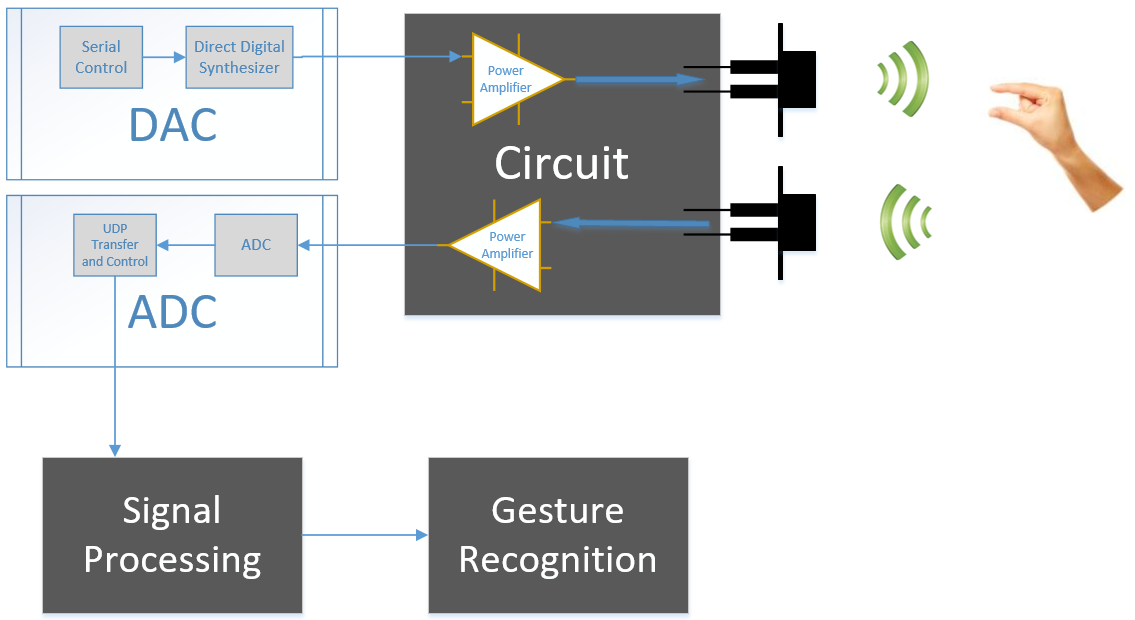}
	\captionsetup{justification=centering}
	\caption{Pipeline Design Block Diagram}	
\end{figure}

\section{Related Work}

Other related work has been performed for micro hand gesture recognition. A project named uTrack utilized magnetic sensing and achieved high finger tracking accuracy but required fingers to be equipped with sensors \cite{chen2013utrack}. Passive infrared (PIR) sensing was exploited for micro hand gestures in the project called Pyro and applicable results were achieved  \cite{gong2017pyro}. By sensing thermal infrared signals radiating from fingers, the researchers implemented signal processing and random forest for thumb-tip gesture recognition. A fine-grained finger gesture recognition system called Mudra \cite{zhang2016mudra} used a WiFi signal to achieved mm-level sensitivity for finger motion. Acoustic sensing also achieved great resolution in finger tracking, which showed potential for micro hand gestures \cite{wang2016device}\cite{nandakumar2016fingerio}.
\begin{figure}[t]
	
	\includegraphics[width=0.5\textwidth]{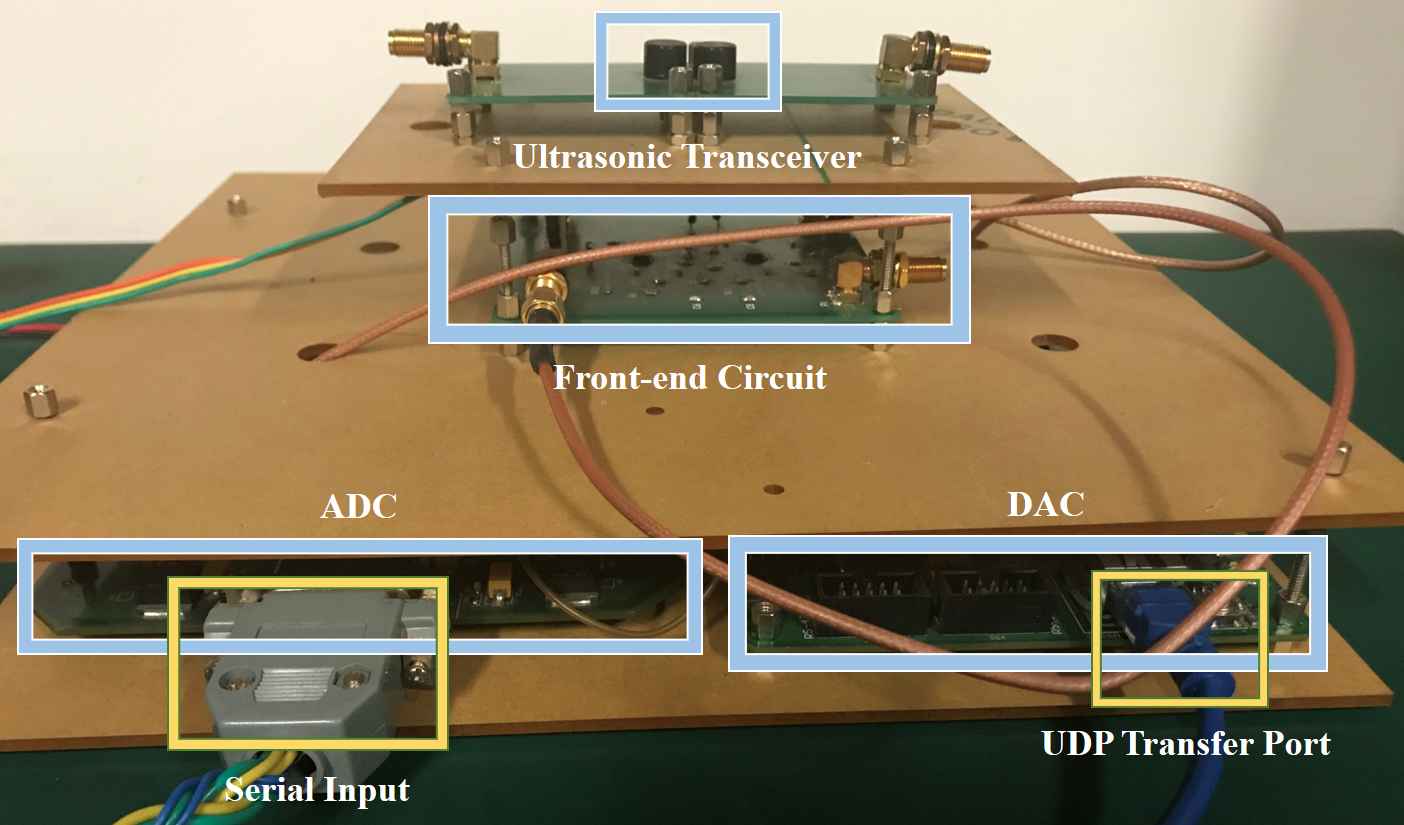}
	\captionsetup{justification=centering}
	\caption{Platform Design}	
\end{figure}

In addition to these works, many radar signal processing technologies combined with machine learning methods or neural network models have been used in hand gesture, and micro hand gesture recognition, especially the Doppler effect. Different kinds of radar, including continuous wave radar \cite{li2017sparsity}, Doppler radar \cite{kim2016hand} and FMCW radar \cite{molchanov2015short}\cite{peng2017fmcw} have been utilized for HGR and achieved results competitive with those of other methods. A K-band continuous wave (CW) radar has been exploited in HGR. It obtained up to 90\% accuracy for 4 wide-range hand gestures \cite{li2017sparsity}. N. Patel et al developed WiSee, a novel gesture recognition system based on a dual-channel Doppler radar that leveraged wireless signals to enable whole-home sensing and recognition under complex conditions \cite{pu2013whole}. These works aroused more interest in utilizing radar signal processing in HGR. However, as we can see, the works above mainly focused on relatively wide-range and large-shift human gestures. As most radar signals' central frequencies and bandwidth limit the feature properties, they are not suitable for recognizing micro hand gestures with subtle motions in several fingers, as they usually cannot distinguish between fingers. The limitations in range and velocity resolution are big challenges for micro hand gesture recognition.

Google announced Project Soli in 2015, which, by utilizing 60-GHz radar, aims at controlling smart devices and wearable devices through micro hand gestures \cite{wang2016interacting}\cite{lien2016soli}. Our project differs from Google Soli in several aspects. We combined the ultrasonic advantage that ultrasonic waves travels at a much slower speed than light and radar signal processing to design our system HUG. Choosing ultrasound waves, we obtain a millimeter-level range and centimeter-per-second-level velocity resolution, much more precise than Soli's millimeter wave radar. Considering the better range-Doppler features, we propose a state transition mechanism to extract and compress features, followed by a HMM model for gesture recognition. This approach achieves competitive results compared to Soli with much concise feature and model sizes of recognition algorithms. Furthermore, to achieve higher accuracy, we implement the computationally more expensive end-to-end network, and achieve a classification accuracy of 96.34\% using our more discriminating range-Doppler features.

Last but not the least, ultrasonic active sensing, particularly with Doppler effects \cite{zhang2007acoustic,zhang2008human,garreau2011gait,dura2011human,raj2012ultrasonic,dura2013multimodal,craley2017action,murray2017bio,kalgaonkar2007ultrasonic}, represents a significant advancement in identification and recognition tasks, especially those involving motion. The design of our ultrasonic radar system draws considerable inspiration from this technology, leveraging its high range and Doppler resolution—features that are highly desirable for micro hand gesture recognition tasks.

\section{Methodology}
\subsection{Platform Design}
To obtain high-quality features, we designed an ultrasonic front-end that transmits ultrasonic waves and receives reflected signals. The parameters of the system are designed to meet our resolution requirement.

To clearly differentiate finger positions and motions, millimeter-level range resolution ($<1$ cm) and centimeter-per-second-level velocity resolution ($<0.1$ m/s) are needed. In addition, the finger velocity is usually no more than $1$ m/s, and the demand for the stop-and-hop assumption has to be met. High precision in range and velocity is also recommended. Considering the radar waveform design and signal processing criteria, the velocity resolution $v_d$ and range resolution $R_d$ are
\begin{eqnarray}
{v_d} = \frac{\lambda }{{2MT}},{R_d} = \frac{c}{{2B}},
\end{eqnarray}
where $c$ is the wave propagation speed, $\lambda$ is the wavelength of the ultrasound, $B$ is the bandwidth, $T$ stands for the pulse repetition interval (PRI), and $M$ is the total number of pulses in a frame.

Thus, to meet the range resolution, we chose the bandwidth $B$ of $20$ kHz. The bandwidth is usually 5\% to 20\% of the central frequency, so we chose ultrasonic waves around the frequency of $300$ kHz \cite{wang2016interacting}\cite{molchanov2015short}. As the maximum finger velocity is $1$ m/s and there is demand for the stop-and-hop assumption, we chose the PRI as $600$ $\upmu$s. Finally, we chose pulse number $M$ as 12 for the requirement of velocity resolution.

Under such parameter settings, the requirements for differentiating fingers' subtle motions are reached, as listed in Table 1. Compared to Soli's millimeter wave radar range resolution of $2.14$ cm, the HUG system is capable of providing much more precise features, which are quite helpful for micro hand gesture recognition.
\begin{table}[htbp]
	\includegraphics[width=0.5\textwidth]{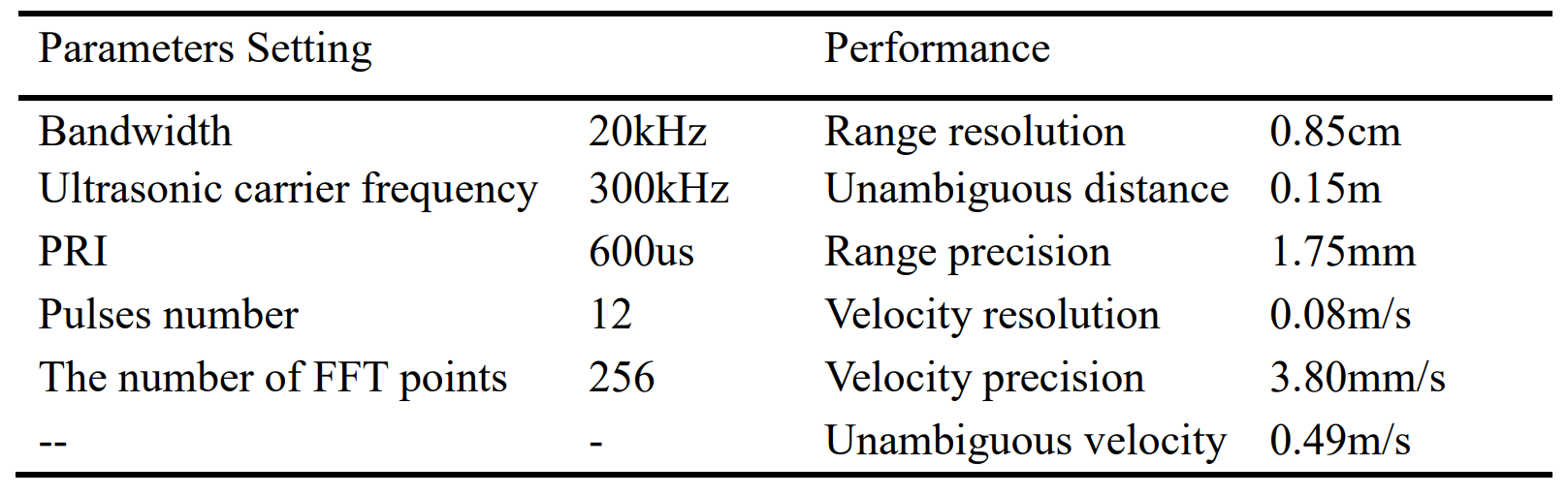}
	\captionsetup{justification=centering}
	\caption{Parameter setting and system performance}	
\end{table}

Fig. 1 illustrates the block diagram of the whole prototype, and the real system is illustrated in Fig. 2. To extract accurate Doppler features, the clock synchronization of DAC and ADC is required. Leakage canceling is implemented in the digital domain. The modulated pulse signal is produced by the DAC and then amplified by the front-end circuit. The amplified signal stimulates an MA300D1-1 Murata Ultrasonic Transceiver to emit ultrasonic waves. When ultrasonic waves are reflected from fingers and palms with noise, received waves are amplified by the front-end circuits before being filled into the ADC. After sampling in ADC, considering the flexibility and real-time capability, the raw signal is transmitted to a PC by Ethernet for the next signal processing module.
\subsection{Range-Doppler Feature Extraction}
In our system, we use the range-Doppler pulsed radar signal processing method to detect the palm and the fingers' movement and to extract features for later classifiers. Pulse radar measures the target range via the round trip delay and utilizes the Doppler effect to determine the velocity by emitting pulses and processing the returned signals at a specific pulse repetition frequency. Two important procedures in the pulse radar signal processing method are fast-time sampling and slow-time sampling. The former refers to the sampling inside the PRI, which determines the range; the latter refers to sampling across multiple pulses using the PRI as the sampling period to determine the Doppler shift for velocity.

To overcome the weakness of echoes and attenuate the influence of noise, chirp pulses are applied to the system as an emitting waveform to improve the SNR while keeping the range resolution. The baseband waveform can be expressed as
\begin{eqnarray}
x(t) = \sum\limits_{m = 0}^{M - 1} {a\left( {t - mT} \right)},
\end{eqnarray}
where\\
\begin{eqnarray}
a(t) = \cos \left( {\pi \frac{B}{\tau }{t^2} - \pi Bt} \right) \qquad 0 \le t \le \tau.
\end{eqnarray}
In the expressions, $a(t)$ is a single pulse with time duration $\tau$ and the other parameters, including $B, T, M$ and PRI, are the same as mentioned above.

The signal processing, after receiving waves are reflected from a hand, is as follows. The received reflected echoes of the ultrasonic waves are processed via the I\&Q quadrature demodulator and the low-pass filter to obtain baseband signals.

\begin{figure}[h]
	\center
	\includegraphics[width=0.3\textwidth]{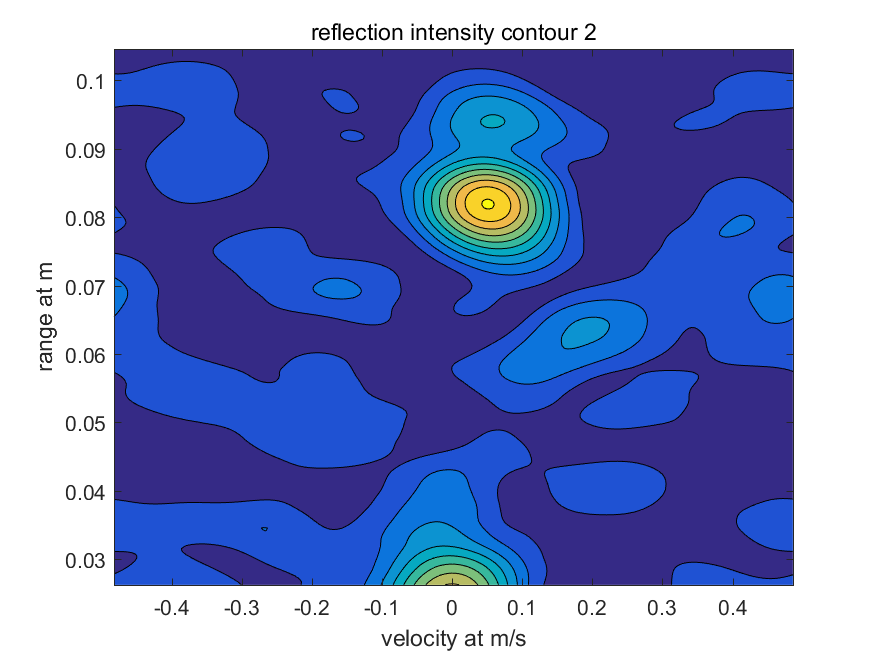}
	\captionsetup{justification=centering}
	\caption{Range-Doppler feature}	
\end{figure}

\begin{figure*}[!t]
	\centering
	\subfloat{\includegraphics[width=0.23\textwidth]{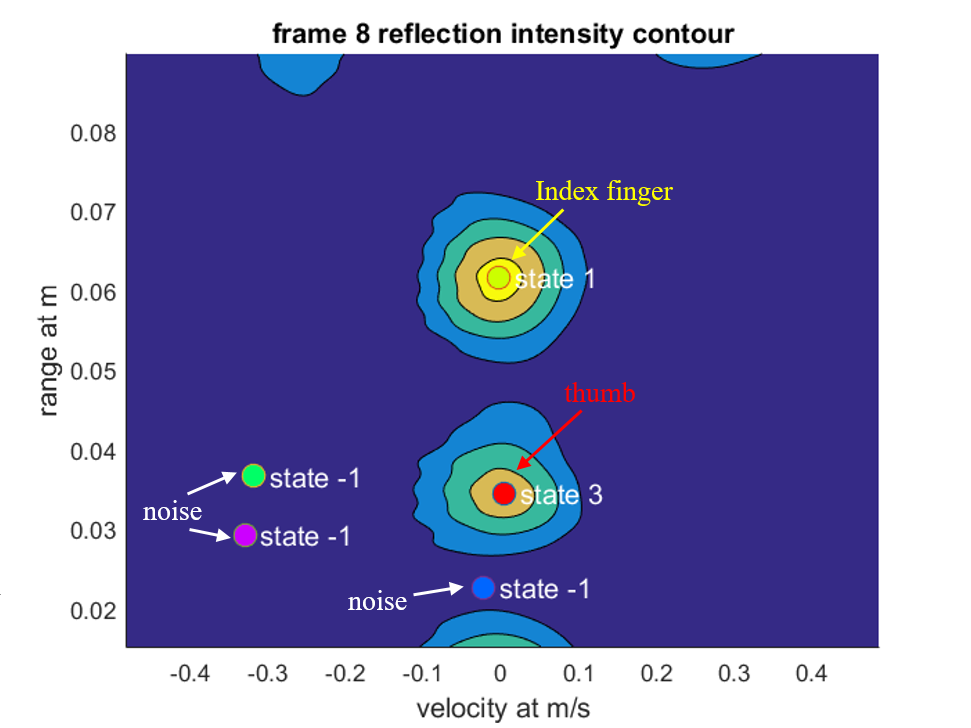}%
		\label{fig_first_case}}
	\hfil
	\subfloat{\includegraphics[width=0.23\textwidth]{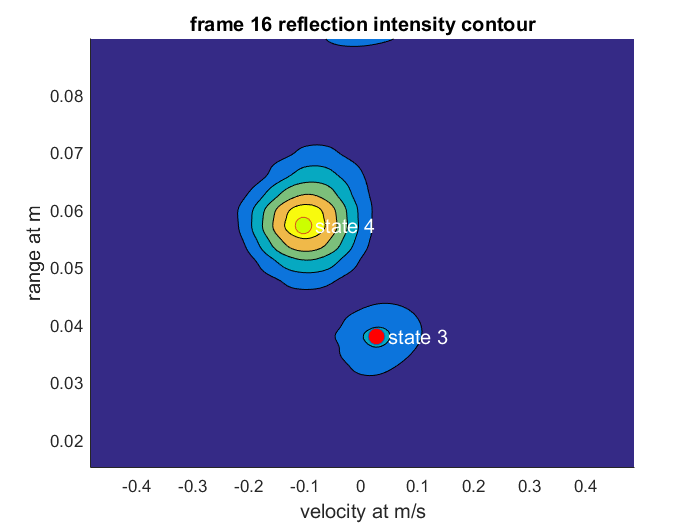}%
		\label{fig_second_case}}
	\hfil
	\subfloat{\includegraphics[width=0.23\textwidth]{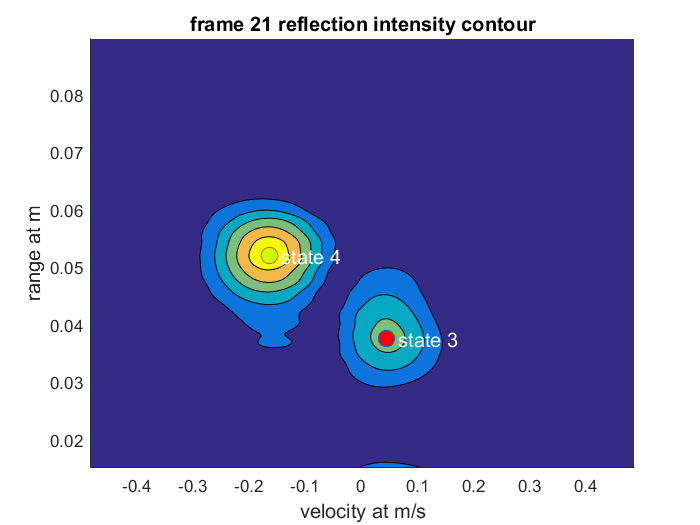}%
		\label{figase}}
	\hfil
	\subfloat{\includegraphics[width=0.23\textwidth]{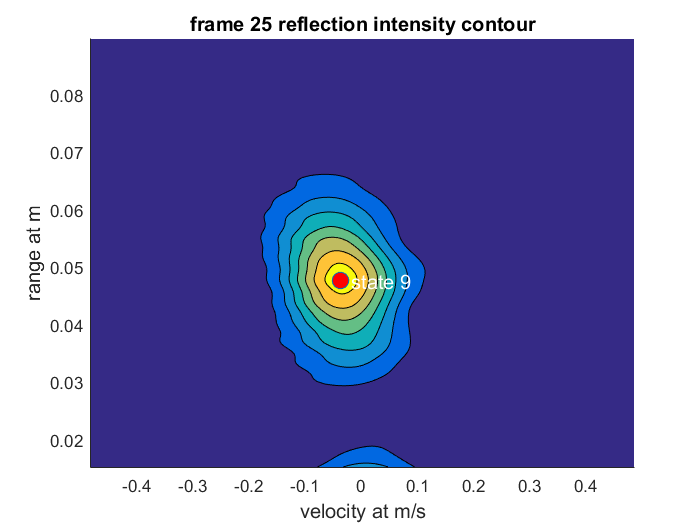}%
		\label{fig_d_case}}
	\caption{Range-Doppler feature images sampled in the ``button off'' (see Fig. \ref{gestures})  gesture. The first subfigure shows the index finger and thumb separated at a 3 cm distance. Noise will be removed using time smoothing to increase robustness to the 3 marked ``noise'' objects in the first subfigure. The second and third subfigures show the index finger moving down with an acceleration while the thumb is nearly motionless with only a tiny upwards velocity. The last subfigure shows two fingers being touched. Note that an object's trajectory will always be a curve in the range-Doppler image. The represented states are labeled at the center of the detected objects.}
	\label{fig_sim}
\end{figure*}

With fast-time sampling at the frequency $F_s$ of $400$ kHz, matched filters are applied to detect the ranges of the palm and fingers. With slow-time sampling, $N$-point FFTs are applied to the samples of different pulses to detect objects' velocities. After two-dimensional sampling, the range-Doppler image features would be generated at a frame level. Every frame of a range-Doppler image is in the shape of $256 \times 180$ as in Fig. 3, which is the region of interest (ROI) of the range-Doppler features at a specific time $m$. The data of a micro hand gesture are a 3-dimensional (range-Doppler-time) cube, which is the stack of range-Doppler images along the time dimension as the output. Fig. 4 shows an example ``button off" gesture, with four range-Doppler images sampled at different times from the data cube.

\subsection{Micro Hand Gesture Recognition}

A pattern recognition module is cascaded after the feature extraction module to classify the time sequential range-Doppler signatures. In our HUG system, we first implemented a state transition-based hidden Markov model, and then, other four different machine learning methods were realized for performance comparison.

\subsubsection{Hidden Markov Model}
Taking advantage of high-resolution range-Doppler images, we can separate contiguous scattering points. We aim at finding a method to extract and compress features containing more efficient information and less noise before feeding them into a classification architecture. Considering the fact that the scattering points of ultrasonic waves correspond to different fingers, the detected peaks in the range-Doppler images can reflect these fingers' motion. For example, the moving trajectory of the scattering peaks shown in Fig. 4 illustrates distinct features of the process of the action ``button off". In addition to precise ranges and velocities of fingers, the different fingers' trajectories and the relations between fingers are exactly the indispensable features we used for recognition. Therefore, we propose a state transition-based HMM for gesture classification. This method consists of three parts cascading, which includes finite-state transition machines, a feature mapping module, and HMM classifiers.
\begin{itemize}
	\item Finite-state Transition Machine
\end{itemize}

A state transition mechanism is proposed to summarize each detected finger's state. The state of one detected scattering point is determined by the following procedure. The reflecting intensity and extent are used to erase noise. In one range-Doppler image, the scattering points' moving direction and velocity determine whether the point is in a static or dynamic condition. Within the range-Doppler images, the distance between the scattering points of adjacent images can be used to determine whether they should be merged with or separated from other points. Tracking mechanisms are used to complement and improve the robustness of the whole state transition processing as well. The detailed design is shown in Fig. 5. Through the extraction process, the features of each frame's range-Doppler image would be represented as an indefinite length vector $\bm{p}_m$. For a gesture, we compress the data to a sequence of vectors
\begin{eqnarray}
\label{equ4}
\mathcal{X}=\left\{ {\bm{p}_m} \right\}_{m = 1}^M,
\end{eqnarray}
where the features of the range-Doppler image in frame $m$ are extracted as a vector $\bm{p}_m$. The vector of symbols $\bm{p}_m$ can be described as (5)
\begin{eqnarray}
\label{equ5}
\bm{p}_m = L\left( {{n_m},{v_m},{r_m}} \right),
\end{eqnarray}
where $n_m$ is the sequence number of scattering points, and $v_m$ and $r_m$ are the velocity and range of the scattering points. $L(\cdot)$  is the state transition function.

\begin{figure}[!hb]
	
	\includegraphics[width=0.5\textwidth]{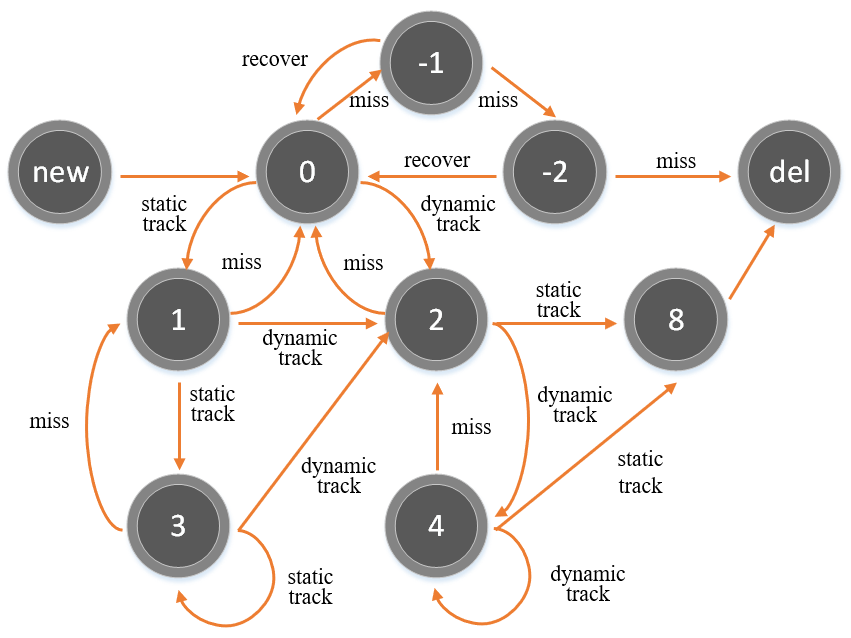}
	\caption{
		State transition diagram. The states represent the types of scattering points, and the arrows denote the methods of transition. Here is an introduction to all the states:
		Initialization states consist of two states, namely, the state `new' and the state `0', where state `new' means the initial states, and state `0' represents uncertain states and the first missing scattering points states. Missing states include states `-1' and `-2', which reflect missing some frames' range-Doppler feature image's scattering points. Tracking states include states `1', `2', `3', and `4', which describe static and dynamic objects' tracking and locking. Ending states include `8' and `del', which denote the end of motion when the dynamic object's speed is less than the threshold or more than 3 frames' scattering points are missing.
	}	
\end{figure}

\begin{itemize}
	\item Feature Mapping Module
\end{itemize}

To fit the features into the HMM model, we represent the features of each range-Doppler image as an integer $s_m$. $S$ is the sequence of range-Doppler images maintaining the time order. The features of each image are extracted to a vector as we mentioned above. A mapping function $\sigma(\cdot)$ is used to discretize and map the vector of integers to the summary variance $s_m$. The lengths of feature vectors are finite, and each element is chosen from a small set of scattering points' states. These variances from each frames form the final feature sequence \[S = \left\{ {{s_m}} \right\}_{m = 1}^M\] defines as (6):
\begin{eqnarray}
{s_m} = \sigma \left( {\bm{p}_m} \right),
\end{eqnarray}
where $\sigma(\cdot)$ is the function which projects the feature vector $\bm{p}_m$ to integer $s_m$.

The feature of each frame is mapped into an integer, which forms this frame's observation of the HMM model. Features in such a form avoid much calculation compared to other methods.
\begin{figure}[!t]
	
	\includegraphics[width=0.5\textwidth]{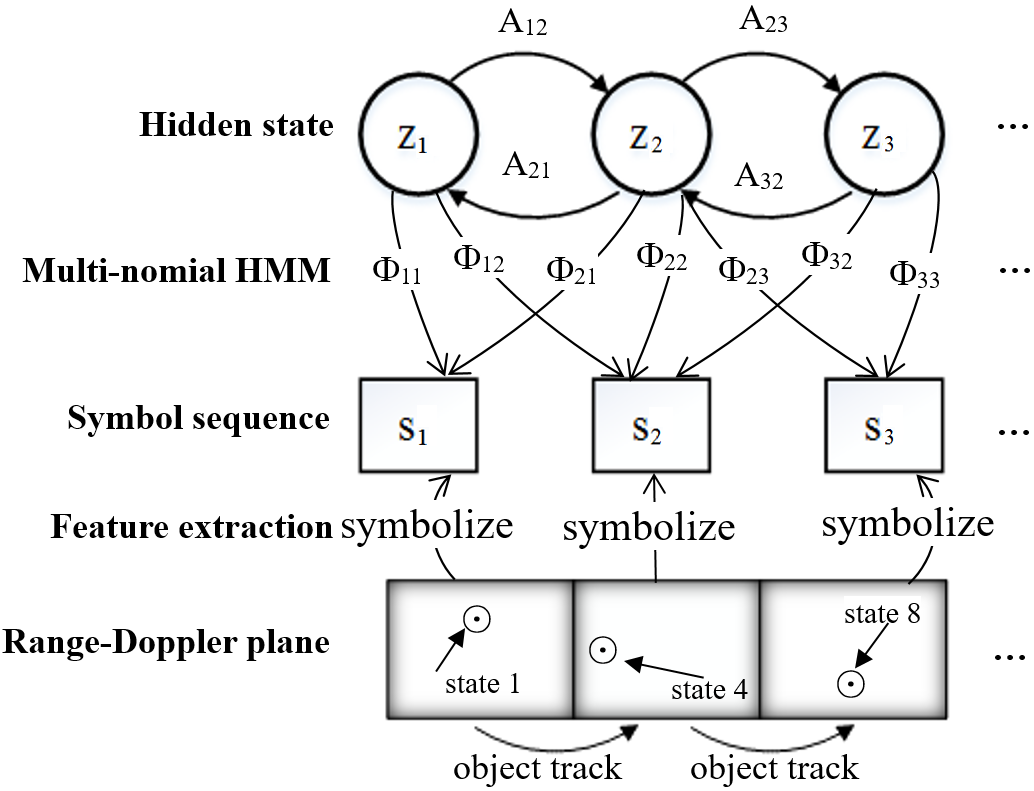}
	\caption{Feature extraction and classification work flow.}
	\label{233}	
\end{figure}
\begin{itemize}
	\item HMM for Classification
\end{itemize}

We use the HMM to extract the patterns of the symbolized states, and Fig. 6 illustrates the work flow. As described in equation (6), the HMM used in this model is composed of a hidden state transition model and a multinomial emission model, where $S = \left\{ {{s_m}} \right\}_{m = 1}^M$ is the observation sequence, $Z = \left\{ {{z_m}} \right\}_{m = 1}^M$ is the hidden state sequence ($z_m$ is the hidden state of time $m$.), and \textbf{A} is the transition probability matrix. $\theta {\rm{ = }}\left\{ {\pi ,\textbf{A},\phi } \right\}$ is the set of all the parameters of the model, where $\pi$ is the initial hidden state distribution and $\phi $ represents the measurement probability. For each pre-defined micro hand gesture, an HMM is trained through the Baum-Welch algorithm \cite{perez2007comparison}. There are $N$ kinds of micro hand gestures, and $C$ is one of them. When the ground truth of an input gesture is class $C$, the likelihood of the models of each kind is calculated through the forward algorithm. Then, we compare all the likelihoods from a Bayesian posterior perspective as described by equation (7):

\[p\left( {S,Z|\theta } \right) = p\left( {{z_1}|\pi } \right)\left[ {\prod\limits_{n = 2}^N {p\left( {{z_n}|{z_{n - 1}},\textbf{A}} \right)} } \right]\prod\limits_{m = 1}^N {p\left( {{s_m}|{z_m},\phi } \right)} ,\]

\begin{eqnarray}
p\left( {C = n|S} \right) = \frac{{p\left( {S|C = n} \right)p\left( {C = n} \right)}}{{\sum\limits_n {p\left( {S|C = n} \right)p\left( {C = n} \right)} }}.
\end{eqnarray}

\subsubsection{Comparison Methods}

For comparison with the state transition-based HMM, we implemented another four frequently used methods, including an end-to-end network, a random forest, a convolutional neural network and a long short-term memory neural network.

Considering that combined networks are widely used and are competitive compared with only CNNs or LSTM networks \cite{vu2016combining}, the end-to-end network is realized and obtains much higher accuracies in hand gesture classification. The detailed network is illustrated in Fig. 7.  We cascade the automatic feature extraction CNN part with the LSTM dynamic gesture classification jointly to form the end-to-end network. In the experiment, we train the two parts together instead of training them individually. We transfer the output of the last convolutional layer of the CNN as a feature sequence of one frame into the LSTM to produce the frame-level classification of hand gestures through a softmax layer. To efficiently use the dynamic process of the micro hand gesture, we add a time-weight filter in a window of 30 frames to obtain gesture-level classification. The filter parameters are empirical by experience. Considering the nearest frame with high weights, the time differential pooling filter was designed as follows:
\[{g_{out}} = \sum\limits_{i = 1}^N {{w_i}\sigma \left[ {g(i)} \right]} , \qquad {w_i} = \frac{{2i}}{N},\] where $g_{out}$ is the gesture prediction, $g(i)$ is the per-frame prediction, $w_i$ are the filter weights.
\begin{figure}[htbp]
	\centering
	\includegraphics[width=0.5\textwidth]{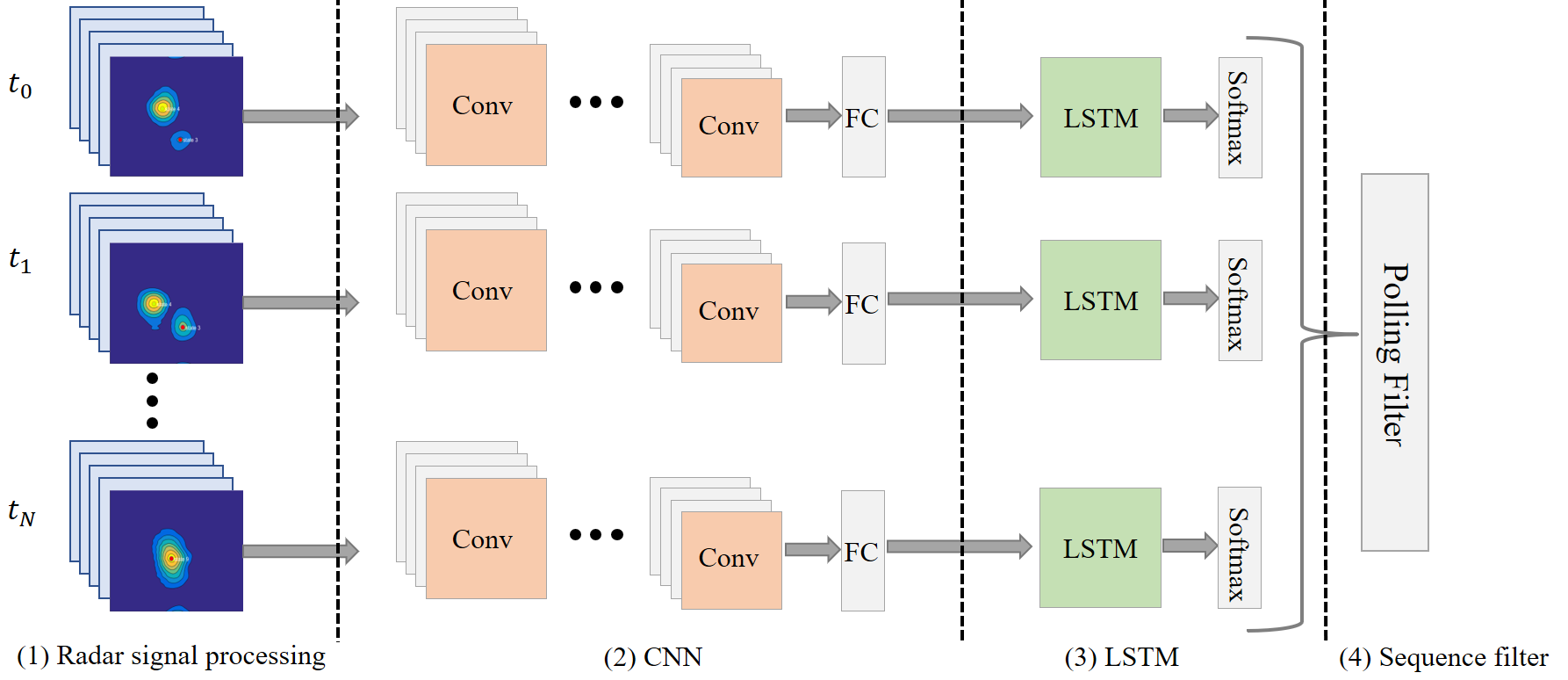}
	\caption{ The end-to-end neural network architecture.}
	\label{end_to_end}
\end{figure}

Another method we implemented is random forest. Random forest has proved to be time and energy efficient in image classification tasks \cite{bosch2007image}. Considering range-Doppler images as regular images, we choose random forest as one approach to implement gesture recognition. A random forest method consisting of 50 trees was realized, and its model size is less than 2 MB.

Among deep learning algorithms, convolutional neural networks are one of the most widely used methods to perform image classification by automatically extracting features through multiple convolutional layers and pooling layers\cite{krizhevsky2012imagenet}. Considering the range-Doppler image cube as a sequence of regular images, the CNN was chosen to be applied to our system. The implemented network consists of five conv-layers and three pooling layers. The concrete architecture is illustrated in Table II. We also implemented the LSTM network, which is a specific kind of recurrent neural network (RNN) suitable for performing time sequential classifications owing to the loops in the network. The network architecture is briefly introduced in Table II.

\section{Implementation and Performance Evaluation}

\subsection{Gesture Set}
There are seven micro hand gestures evaluated in our system. As shown in Fig. 8, we have six gestures (indexed from the 2nd to the 7th), including finger press, button on, button down, motion up, motion down, and screw. Other kinds of gestures, including the case with no finger, are categorized into the 1st kind, i.e., no-finger. The principles used in our design are as follows: (1) The micro hand gestures chosen only involve subtle finger motions; the motions of large muscle groups in the wrist and arm are not included. (2) All gestures are dynamic gestures, which enable ultrasonic waves to capture Doppler features of the fingers' motions.
\begin{figure}[htbp]
	\centering
	\includegraphics[width=0.5\textwidth]{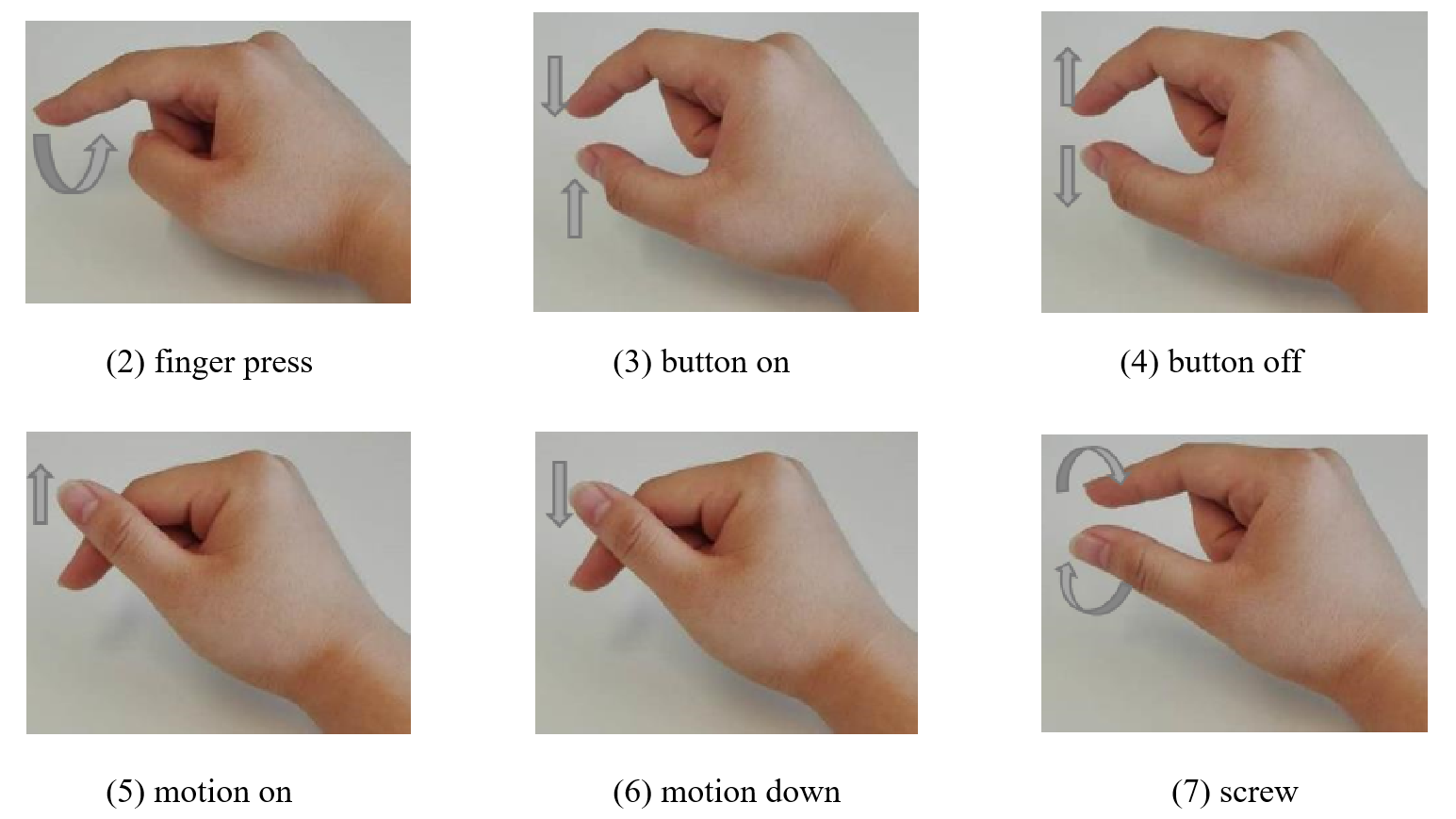}
	\caption{Examples of micro hand gestures, named finger press, button on, button off, motion up, motion down, and screw. The 1st one is no-finger.}
	\label{gestures}
\end{figure}
\subsection{Data Acquisition}
To ensure that adequate variation of gesture execution resulted from individual characteristics and other environmental facets, we demanded that subjects repeat all the gestures. We invited nine subjects to execute the six designed gestures, only giving sketchy instructions for how to perform these gestures. Each kind of gesture was labeled, without removing any data from the dataset to ensure large variance. For each gesture, 50 samples were recorded from each subject, resulting in $9 \times 6 \times 50 = 2700$ sample sequences. In addition to these 6 gestures, we added no-finger as another kind of gesture (labeled as 0). We obtained another $2700$ samples for the no-finger, and these samples were divided into each subjects' sub-datasets.
Finally, a dataset of 5400 samples in total was used in all of our experiments as the raw input.

For training and validation, we used standard $k$-fold leave-one-subject-out, where $k$ is the nine subjects in our experiments.

\subsection{Experimental Implementation}
The implementation details of all the five methods are shown in Table II.

For our particular HMM architecture, as it specializes in dealing with input of unfixed length, we just feed sequences of the features with different lengths into models and train seven models for the seven micro gestures respectively. After extracting frame-level features using the state transition machine, we embedded these features into the final feature sequence, $S$, according to the produced dictionary. As expected, the training process cost less time compared to those of neural network methods. In our experiments, only ten iteration loops were used. The classification is based on the maximum a posteriori method, where the prior probability is used to adjust the classification owing to no-finger being the usual case.

For the random forest method, we use a forest size of 50 with no maximum depth. We downsampled the time sequence feature cube to the tensor size of $64 \times 45 \times 30$ (height, width, time sequence) and transformed it into a one-dimensional sequence. Then these one-dimensional feature sequences were fed into the random forest for classification.

The three deep neural network methods, the end-to-end, the CNN, and the LSTM, have similar training procedures. For the end-to-end and the CNN, as the gestures sampled in our dataset are of variable length, mainly from 30 frames to 120 frames, we downsampled the raw data to a fixed length of $256 \times 180 \times 30$ as a three-dimensional input. For the LSTM, to reduce the feature size and obtain more data, we divide each raw sequence data into a uniform length of 30 frames and then subsample them into a size of $128 \times 90 \times 30$. Then the produced samples are stochastically shuffled in the training procedure, and the time order is maintained within each sequence. The alignment of samples makes mini-batch training possible and accelerates the training process. To avoid local optima and dampening oscillation, and also for the training efficiency, we use mini-batch adaptive moment estimation (Adam) \cite{kingma2014adam} with the initial learning rate of 0.001 and a batch size of 16 or 32 for all the neural networks. In the training process of the end-to-end, the fully connected layer after the convolutional layers in the CNN is connected to the LSTM network directly and trained jointly.

\begin{table*}[!t]
	\centering
	\includegraphics[width=1.0\textwidth]{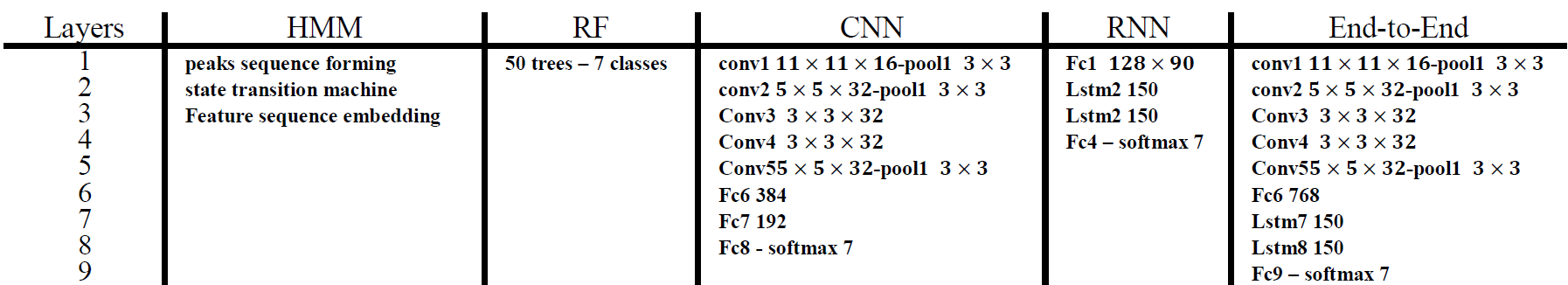}
	\caption{Network architecture used in our experiments}
	\label{model_architecture}
\end{table*}

\begin{table*}[!t]
	\centering	
	\includegraphics[width=1.0\textwidth]{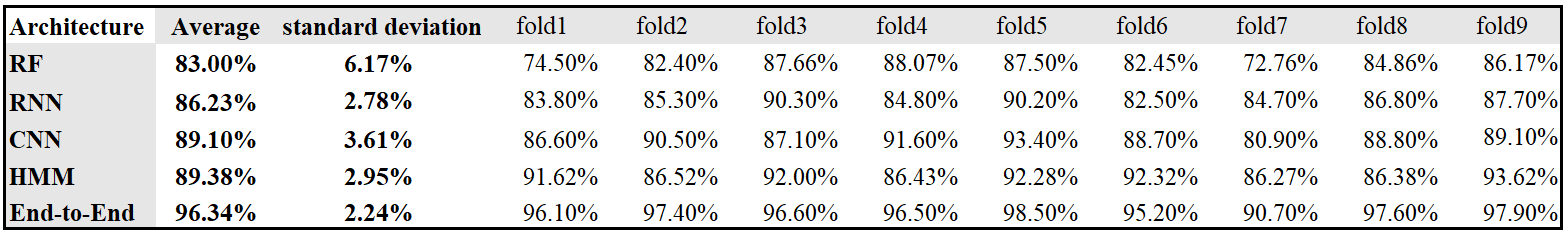}
	\caption{Accuracies of various classification methods for 9 subjects and their average fold$1{\rm{ - }}$fold$9$ representing the 9 subjects.}
	\label{9fold_result}
\end{table*}

\subsection{Computation Complexity and Real-time Performance}
All the training and tests were carried out on the platform with an Nvidia Tesla K20 GPU. The five methods described above all achieved real-time micro hand gesture recognition in our system. Among them, the end-to-end architecture has the biggest model size of 150 MB and the model size of random forest, standalone CNN and RNN are all on the order of MB. However, our proposed HMM-based method is just 40 KB, which is much smaller than the others. For real-time performance, considering frame-level prediction, the end-to-end processed 450 frames per second, whereas the Soli's end-to-end predicts frames at 150 Hz \cite{wang2016interacting}. Furthermore, for gesture level prediction, our proposed state transition-based HMM predicts gestures at a much higher rate (250 Hz) than the end-to-end neural network method (15 Hz). On account of the small size and high prediction rate, the state-transition-based HMM has great potential to be embedded into wearable and portable devices.

\subsection{Results and Performance Evaluation}
For comparison of micro hand gesture classification, we select equal numbers of relatively micro hand gestures from the 11 hand gestures that Google's Soli used in \cite{wang2016interacting}. Excluding four wide-range hand gestures, there are actually six relatively micro hand gestures, including pinch index, pinch pinky, finger slide, finger rub, slow swipe, and fast swipe, as well as a static gesture (palm hold), which are relevant to our work \cite{gong2017pyro}. For comparison, we delete the rest four kinds of wide-range hand gesture results and calculated the confusion matrix again. Because the micro hand gestures are not easy to be incorrectly recognized as wide-range gestures, we transfer the corresponding false negatives as true positives when calculating the new confusion matrix. According to the positive calculation, the average classification accuracy of the seven micro hand gestures is approximately 88\%.

Fig. 9 shows the confusion matrix of the results achieved by our state-transition-based HMM model. The average of the accuracy is 89\%, which is competitive compared to Soli but obtains a balance between accuracy and computation load. Due to the model's small size and high prediction rate, the state-transition-based HMM model has great potential to be used in wearable and embedded systems such as smart watches or virtual reality.
\begin{figure}[htbp]
	\centering
	\includegraphics[width=0.5\textwidth]{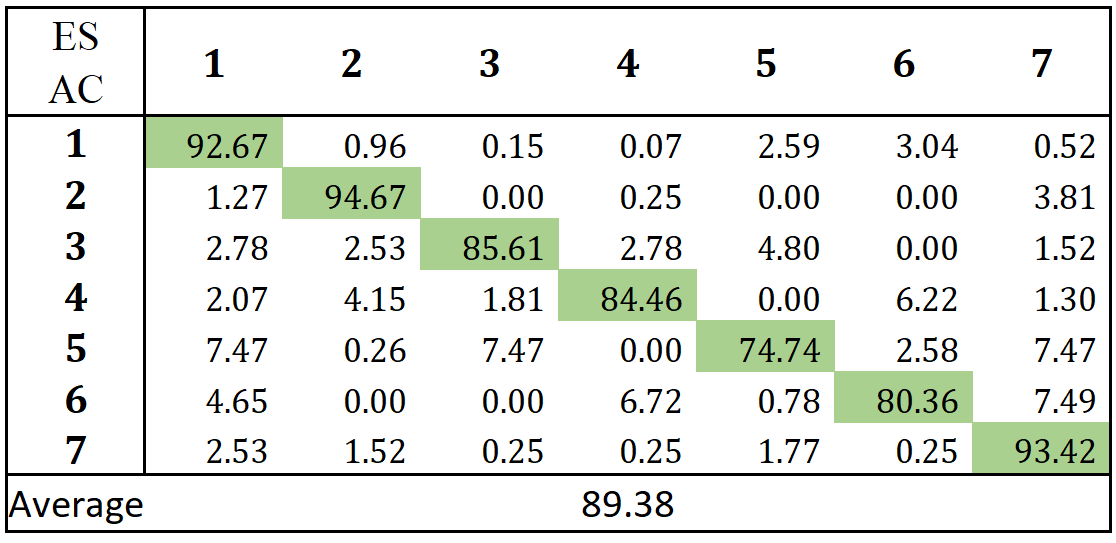}
	\caption{Confusion matrix of accuracy of the HMM.}
	\label{hmm_result}
\end{figure}

Fig. 10 shows the confusion matrix corresponding to the end-to-end network, whose average accuracy is as high as 96.34\%. Profiting from the advantages of the ultrasonic signal's high resolution, the end-to-end architecture can achieve highly accurate classification, where the highest accuracy is 98.89\%, and the lowest is 89.47\%.

\begin{figure}[htbp]
	\centering	\includegraphics[width=0.5\textwidth]{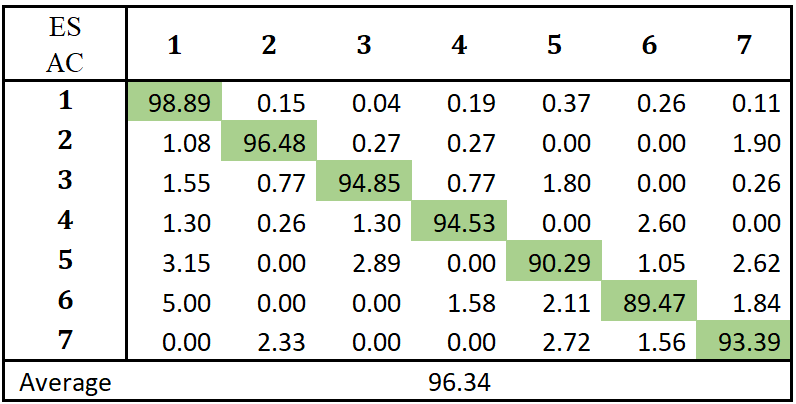}
	\caption{Confusion matrix of accuracy of the end-to-end.}
	\label{end2end_result}
\end{figure}

Table III illustrates the various subject classification validation accuracies of all the five classification methods, using leave-one-subject-out cross-validation. Owing to the personal variance in gesture execution, we use 8 subjects' samples and leave the last subject's samples separately for validation. The experiment can be a good predictor of real-world gesture recognition, as the validation datasets are entirely separated from the training sets. The variance between the subjects shown in Table III, which is less than 7\% in these methods, also verifies that our architecture has good generalization capability.

\begin{table}[htbp]
	\centering
	\includegraphics[width=0.5\textwidth]{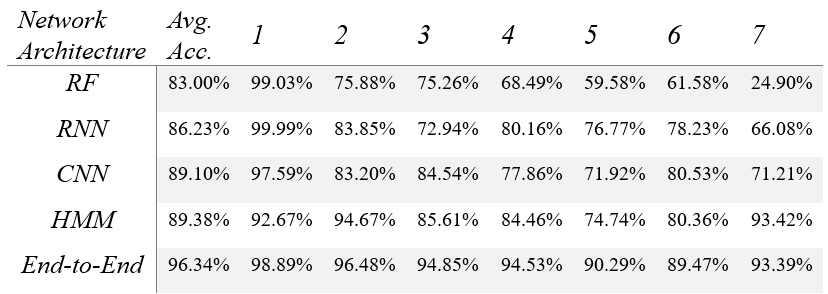}
	\caption{The accuracy of each gesture of different classification methods. All of them use leave-one-subject-out cross-validation: (1) random forest with per-gesture sequence accuracy, (2) CNN with per-gesture sequence accuracy, (3) RNN with sequence pooling of per-frame prediction, and (4) end-to-end with sequence pooling of per-frame prediction}
	\label{gesture_result}
\end{table}
Table IV introduces the overall classification accuracy of different architectures. We use sequence-level recognition instead of frame-level classification because our micro hand gestures are all dynamic ones. As we can see, for the 7th gesture, the classification accuracies of the state-transition-based HMM and the end-to-end are much higher than other three methods. The 1st gesture, which is labeled as no-finger, achieves the highest accuracy among all the classification methods. The classification accuracy of the 5th gesture is the worst among seven gestures by all the five methods. The reason is that this gesture is easy to be classified into the 3rd gesture owing to their similar features.

A demonstration of HUG system is presented in \cite{hug.org}, where a prototype was realized to control a music player by micro hand gestures in real-time. In this demo, one can find that our system HUG can recognize 7 kinds of gestures and realize the interaction between the user and the music player.

\section{Conclusion}
In this paper, we proposed a system and methods for micro hand gesture recognition by using ultrasonic active sensing. Owing to the high resolution provided by ultrasonic waves, better quality features can be extracted. The proposed state-transition-based HMM method, which has less computational complexity, achieved a comparable 89.38\% classification accuracy. Furthermore, we used an end-to-end method and achieved a classification accuracy of 96.34\%. Our system and methods of recognizing micro hand gestures showed great potential in HCI, especially in scenarios where a touchless input modality is more attractive, such as for wearable and portable devices or for driving assistance.



\section*{Acknowledgment}

We are grateful to all the participants for their efforts in collecting hand gesture data for our experiments. This work is supported by the National Natural Science Foundation of China (Grant No. 61571260 and No. 61801258).

\ifCLASSOPTIONcaptionsoff
\newpage
\fi



%

\bibliographystyle{IEEEtran}  
\bibliography{IEEEabrv,hug}

\begin{thebibliography}{10}
\providecommand{\url}[1]{#1}
\csname url@samestyle\endcsname
\providecommand{\newblock}{\relax}
\providecommand{\bibinfo}[2]{#2}
\providecommand{\BIBentrySTDinterwordspacing}{\spaceskip=0pt\relax}
\providecommand{\BIBentryALTinterwordstretchfactor}{4}
\providecommand{\BIBentryALTinterwordspacing}{\spaceskip=\fontdimen2\font plus
\BIBentryALTinterwordstretchfactor\fontdimen3\font minus
  \fontdimen4\font\relax}
\providecommand{\BIBforeignlanguage}[2]{{%
\expandafter\ifx\csname l@#1\endcsname\relax
\typeout{** WARNING: IEEEtran.bst: No hyphenation pattern has been}%
\typeout{** loaded for the language `#1'. Using the pattern for}%
\typeout{** the default language instead.}%
\else
\language=\csname l@#1\endcsname
\fi
#2}}
\providecommand{\BIBdecl}{\relax}
\BIBdecl

\bibitem{kjeldskov2003review}
J.~Kjeldskov and C.~Graham, ``A review of mobile {HCI} research methods,'' in
  \emph{International Conference on Mobile Human-Computer Interaction}.\hskip
  1em plus 0.5em minus 0.4em\relax Springer, 2003, pp. 317--335.

\bibitem{pickering2007research}
C.~A. Pickering, K.~J. Burnham, and M.~J. Richardson, ``A research study of
  hand gesture recognition technologies and applications for human vehicle
  interaction,'' in \emph{Automotive Electronics, 2007 3rd Institution of
  Engineering and Technology Conference on}.\hskip 1em plus 0.5em minus
  0.4em\relax IET, 2007, pp. 1--15.

\bibitem{pu2013whole}
Q.~Pu, S.~Gupta, S.~Gollakota, and S.~Patel, ``Whole-home gesture recognition
  using wireless signals,'' in \emph{Proceedings of the 19th annual
  international conference on Mobile computing \& networking}.\hskip 1em plus
  0.5em minus 0.4em\relax ACM, 2013, pp. 27--38.

\bibitem{gong2017pyro}
J.~Gong, Y.~Zhang, X.~Zhou, and X.-D. Yang, ``Pyro: Thumb-tip gesture
  recognition using pyroelectric infrared sensing,'' in \emph{Proceedings of
  the 30th Annual ACM Symposium on User Interface Software and
  Technology}.\hskip 1em plus 0.5em minus 0.4em\relax ACM, 2017, pp. 553--563.

\bibitem{fingo.org}
{uSens}, ``Fingo,'' \url{https://www.usens.com/fingo}.

\bibitem{ye2014current}
H.~Ye, M.~Malu, U.~Oh, and L.~Findlater, ``Current and future mobile and
  wearable device use by people with visual impairments,'' in \emph{Proceedings
  of the SIGCHI Conference on Human Factors in Computing Systems}.\hskip 1em
  plus 0.5em minus 0.4em\relax ACM, 2014, pp. 3123--3132.

\bibitem{wang2016interacting}
S.~Wang, J.~Song, J.~Lien, I.~Poupyrev, and O.~Hilliges, ``Interacting with
  {Soli}: Exploring fine-grained dynamic gesture recognition in the
  radio-frequency spectrum,'' in \emph{Proceedings of the 29th Annual Symposium
  on User Interface Software and Technology}.\hskip 1em plus 0.5em minus
  0.4em\relax ACM, 2016, pp. 851--860.

\bibitem{chen2003hand}
F.-S. Chen, C.-M. Fu, and C.-L. Huang, ``Hand gesture recognition using a
  real-time tracking method and hidden {M}arkov models,'' \emph{Image and
  vision computing}, vol.~21, no.~8, pp. 745--758, 2003.

\bibitem{yao2015hand}
Y.~Yao and C.-T. Li, ``Hand gesture recognition and spotting in uncontrolled
  environments based on classifier weighting,'' in \emph{Image Processing
  (ICIP), 2015 IEEE International Conference on}.\hskip 1em plus 0.5em minus
  0.4em\relax IEEE, 2015, pp. 3082--3086.

\bibitem{ren2011depth}
Z.~Ren, J.~Meng, and J.~Yuan, ``Depth camera based hand gesture recognition and
  its applications in human-computer-interaction,'' in \emph{Information,
  Communications and Signal Processing (ICICS) 2011 8th International
  Conference on}.\hskip 1em plus 0.5em minus 0.4em\relax IEEE, 2011, pp. 1--5.

\bibitem{suarez2012hand}
J.~Suarez and R.~R. Murphy, ``Hand gesture recognition with depth images: A
  review,'' in \emph{Ro-man, 2012 IEEE}.\hskip 1em plus 0.5em minus 0.4em\relax
  IEEE, 2012, pp. 411--417.

\bibitem{song2014air}
J.~Song, G.~S{\"o}r{\"o}s, F.~Pece, S.~R. Fanello, S.~Izadi, C.~Keskin, and
  O.~Hilliges, ``In-air gestures around unmodified mobile devices,'' in
  \emph{Proceedings of the 27th annual ACM symposium on User interface software
  and technology}.\hskip 1em plus 0.5em minus 0.4em\relax ACM, 2014, pp.
  319--329.

\bibitem{zhang2016mudra}
O.~Zhang and K.~Srinivasan, ``Mudra: User-friendly {F}ine-grained {G}esture
  {R}ecognition using {WiFi} {S}ignals,'' in \emph{Proceedings of the 12th
  International on Conference on emerging Networking EXperiments and
  Technologies}.\hskip 1em plus 0.5em minus 0.4em\relax ACM, 2016, pp. 83--96.

\bibitem{gupta2012soundwave}
S.~Gupta, D.~Morris, S.~Patel, and D.~Tan, ``Soundwave: using the {D}oppler
  effect to sense gestures,'' in \emph{Proceedings of the SIGCHI Conference on
  Human Factors in Computing Systems}.\hskip 1em plus 0.5em minus 0.4em\relax
  ACM, 2012, pp. 1911--1914.

\bibitem{chen2013utrack}
K.-Y. Chen, K.~Lyons, S.~White, and S.~Patel, ``u{T}rack: {3D} input using two
  magnetic sensors,'' in \emph{Proceedings of the 26th annual ACM symposium on
  User interface software and technology}.\hskip 1em plus 0.5em minus
  0.4em\relax ACM, 2013, pp. 237--244.

\bibitem{li2017sparsity}
G.~Li, R.~Zhang, M.~Ritchie, and H.~Griffiths, ``Sparsity-based dynamic hand
  gesture recognition using micro-{D}oppler signatures,'' in \emph{Radar
  Conference (RadarConf), 2017 IEEE}.\hskip 1em plus 0.5em minus 0.4em\relax
  IEEE, 2017, pp. 0928--0931.

\bibitem{molchanov2015short}
P.~Molchanov, S.~Gupta, K.~Kim, and K.~Pulli, ``Short-range {FMCW} monopulse
  radar for hand-gesture sensing,'' in \emph{Radar Conference (RadarCon), 2015
  IEEE}.\hskip 1em plus 0.5em minus 0.4em\relax IEEE, 2015, pp. 1491--1496.

\bibitem{skolnik1962introduction}
M.~I. Skolnik, ``Introduction to radar,'' \emph{Radar Handbook}, vol.~2, 1962.

\bibitem{wang2016device}
W.~Wang, A.~X. Liu, and K.~Sun, ``Device-free gesture tracking using acoustic
  signals,'' in \emph{Proceedings of the 22nd Annual International Conference
  on Mobile Computing and Networking}.\hskip 1em plus 0.5em minus 0.4em\relax
  ACM, 2016, pp. 82--94.

\bibitem{nandakumar2016fingerio}
R.~Nandakumar, V.~Iyer, D.~Tan, and S.~Gollakota, ``Fingerio: Using active
  sonar for fine-grained finger tracking,'' in \emph{Proceedings of the 2016
  CHI Conference on Human Factors in Computing Systems}.\hskip 1em plus 0.5em
  minus 0.4em\relax ACM, 2016, pp. 1515--1525.

\bibitem{timmurphy.org}
{Snapdragon}, ``Snapdragon sense {ID},''
  \url{https://www.qualcomm.com/products/features/security/fingerprint-sensors}.

\bibitem{przybyla20153d}
R.~J. Przybyla, H.-Y. Tang, A.~Guedes, S.~E. Shelton, D.~A. Horsley, and B.~E.
  Boser, ``{3D} ultrasonic rangefinder on a chip,'' \emph{IEEE Journal of
  Solid-State Circuits}, vol.~50, no.~1, pp. 320--334, 2015.

\bibitem{godrich2017students}
H.~Godrich, ``Students' design project series: Sharing experiences [sp
  education],'' \emph{IEEE Signal Processing Magazine}, vol.~34, no.~1, pp.
  82--88, 2017.

\bibitem{kim2016hand}
Y.~Kim and B.~Toomajian, ``Hand gesture recognition using micro-{D}oppler
  signatures with convolutional neural network,'' \emph{IEEE Access}, vol.~4,
  pp. 7125--7130, 2016.

\bibitem{peng2017fmcw}
Z.~Peng, C.~Li, J.-M. Mu{\~n}oz-Ferreras, and R.~G{\'o}mez-Garc{\'\i}a, ``An
  {FMCW} radar sensor for human gesture recognition in the presence of multiple
  targets,'' in \emph{Microwave Bio Conference (IMBIOC), 2017 First IEEE MTT-S
  International}.\hskip 1em plus 0.5em minus 0.4em\relax IEEE, 2017, pp. 1--3.

\bibitem{lien2016soli}
J.~Lien, N.~Gillian, M.~E. Karagozler, P.~Amihood, C.~Schwesig, E.~Olson,
  H.~Raja, and I.~Poupyrev, ``Soli: Ubiquitous gesture sensing with millimeter
  wave radar,'' \emph{ACM Transactions on Graphics (TOG)}, vol.~35, no.~4, p.
  142, 2016.

\bibitem{zhang2007acoustic}
Z.~Zhang, P.~O. Pouliquen, A.~Waxman, and A.~G. Andreou, ``Acoustic
  micro-doppler radar for human gait imaging,'' \emph{The Journal of the
  Acoustical Society of America}, vol. 121, no.~3, pp. EL110--EL113, 2007.

\bibitem{zhang2008human}
Z.~Zhang and A.~G. Andreou, ``Human identification experiments using acoustic
  micro-doppler signatures,'' in \emph{2008 Argentine School of
  Micro-Nanoelectronics, Technology and Applications}.\hskip 1em plus 0.5em
  minus 0.4em\relax IEEE, 2008, pp. 81--86.

\bibitem{garreau2011gait}
G.~Garreau, C.~M. Andreou, A.~G. Andreou, J.~Georgiou, S.~Dura-Bernal,
  T.~Wennekers, and S.~Denham, ``Gait-based person and gender recognition using
  micro-doppler signatures,'' in \emph{2011 IEEE Biomedical Circuits and
  Systems Conference (BioCAS)}.\hskip 1em plus 0.5em minus 0.4em\relax IEEE,
  2011, pp. 444--447.

\bibitem{dura2011human}
S.~Dura-Bernal, G.~Garreau, C.~Andreou, A.~Andreou, J.~Georgiou, T.~Wennekers,
  and S.~Denham, ``Human action categorization using ultrasound micro-doppler
  signatures,'' in \emph{Human Behavior Understanding: Second International
  Workshop, HBU 2011, Amsterdam, The Netherlands, November 16, 2011.
  Proceedings 2}.\hskip 1em plus 0.5em minus 0.4em\relax Springer, 2011, pp.
  18--28.

\bibitem{raj2012ultrasonic}
B.~Raj, K.~Kalgaonkar, C.~Harrison, and P.~Dietz, ``Ultrasonic doppler sensing
  in hci,'' \emph{IEEE Pervasive Computing}, vol.~11, no.~2, pp. 24--29, 2012.

\bibitem{dura2013multimodal}
S.~Dura-Bernal, G.~Garreau, J.~Georgiou, A.~G. Andreou, S.~L. Denham, and
  T.~Wennekers, ``Multimodal integration of micro-doppler sonar and auditory
  signals for behavior classification with convolutional networks,''
  \emph{International journal of neural systems}, vol.~23, no.~05, p. 1350021,
  2013.

\bibitem{craley2017action}
J.~Craley, T.~S. Murray, D.~R. Mendat, and A.~G. Andreou, ``Action recognition
  using micro-doppler signatures and a recurrent neural network,'' in
  \emph{2017 51st Annual Conference on Information Sciences and Systems
  (CISS)}.\hskip 1em plus 0.5em minus 0.4em\relax IEEE, 2017, pp. 1--5.

\bibitem{murray2017bio}
T.~S. Murray, D.~R. Mendat, K.~A. Sanni, P.~O. Pouliquen, and A.~G. Andreou,
  ``Bio-inspired human action recognition with a micro-doppler sonar system,''
  \emph{IEEE Access}, vol.~6, pp. 28\,388--28\,403, 2017.

\bibitem{kalgaonkar2007ultrasonic}
K.~Kalgaonkar, R.~Hu, and B.~Raj, ``Ultrasonic doppler sensor for voice
  activity detection,'' \emph{IEEE Signal Processing Letters}, vol.~14, no.~10,
  pp. 754--757, 2007.

\bibitem{perez2007comparison}
{\'O}.~P{\'e}rez, M.~Piccardi, J.~Garc{\'\i}a, M.~{\'A}. Patricio, and J.~M.
  Molina, ``Comparison between genetic algorithms and the baum-welch algorithm
  in learning {HMMs} for human activity classification,'' in \emph{Workshops on
  Applications of Evolutionary Computation}.\hskip 1em plus 0.5em minus
  0.4em\relax Springer, 2007, pp. 399--406.

\bibitem{vu2016combining}
N.~T. Vu, H.~Adel, P.~Gupta, and H.~Sch{\"u}tze, ``Combining recurrent and
  convolutional neural networks for relation classification,'' \emph{arXiv
  preprint arXiv:1605.07333}, 2016.

\bibitem{bosch2007image}
A.~Bosch, A.~Zisserman, and X.~Munoz, ``Image classification using random
  forests and ferns,'' in \emph{Computer Vision, 2007. ICCV 2007. IEEE 11th
  International Conference on}.\hskip 1em plus 0.5em minus 0.4em\relax IEEE,
  2007, pp. 1--8.

\bibitem{krizhevsky2012imagenet}
A.~Krizhevsky, I.~Sutskever, and G.~E. Hinton, ``Imagenet classification with
  deep convolutional neural networks,'' in \emph{Advances in neural information
  processing systems}, 2012, pp. 1097--1105.

\bibitem{kingma2014adam}
D.~P. Kingma and J.~Ba, ``Adam: A method for stochastic optimization,''
  \emph{arXiv preprint arXiv:1412.6980}, 2014.

\bibitem{hug.org}
HUG, ``Hug music player controlling,''
  \url{https://www.youtube.com/watch?v=8FgdiIb9WqY}.

\end{thebibliography}


\begin{thebibliography}{1}
	
	\bibitem{ref21}
	J Kjeldskov, C Graham,\emph{A Review of Mobile HCI Research Methods}.\hskip 1em plus
	0.5em minus 0.4em\relax Springer Berlin Heidelberg, 2003.
	
	\bibitem{ref22}
	Pickering C A, Burnham K J, Richardson M J,\emph{A Research Study of Hand Gesture Recognition Technologies and Applications for Human Vehicle Interaction}.\hskip 1em plus
	0.5em minus 0.4em\relax Automotive Electronics, 2007, Institution of Engineering and Technology Conference on. IET, 2007:1-15.
	
	\bibitem{ref11}
	Pu, Qifan and Gupta, Sidhant and Gollakota, Shyamnath and Patel, Shwetak, \emph{Whole-home gesture recognition using wireless signals}.\hskip 1em plus
	0.5em minus 0.4em\relax Proceedings of the 19th annual international conference on Mobile computing \& networking, 2013.
	
	\bibitem{ref23}
	Gong J, Zhang Y, Zhou X, et al,\emph{Pyro: Thumb-Tip Gesture Recognition Using Pyroelectric Infrared Sensing}.\hskip 1em plus
	0.5em minus 0.4em\relax he, ACM Symposium. ACM, 2017
	
	\bibitem{ref24}
	uSens,\emph{Fingo}.\hskip 1em plus
	0.5em minus 0.4em\relax
	
	\bibitem{ref25}
	Ye, Hanlu and Malu, Meethu and Oh, Uran and Findlater, Leah,\emph{Current and future mobile and wearable device use by people with visual impairments}.\hskip 1em plus
	0.5em minus 0.4em\relax ACM, 2014
	
	\bibitem{ref26}
	Chen K Y, Lyons K, White S,\emph{uTrack: 3D input using two magnetic sensors.}.\hskip 1em plus
	0.5em minus 0.4em\relax Proceedings of the 3rd International Conference on Frontiers of Intelligent Computing: Theory and Applications (FICTA) 2014. Springer International Publishing, 2015:165-174.
	
	\bibitem{ref27}
	Wang W, Liu A X, Sun K,\emph{Device-free gesture tracking using acoustic signals}.\hskip 1em plus
	0.5em minus 0.4em\relax International Conference on Mobile Computing and NETWORKING. ACM, 2016
	
	\bibitem{ref28}
	Gollakota, Shyamnath and Gollakota, Shyamnath and Gollakota, Shyamnath and Gollakota, Shyamnath,\emph{FingerIO: Using Active Sonar for Fine-Grained Finger Tracking}.\hskip 1em plus
	0.5em minus 0.4em\relax CHI Conference on Human Factors in Computing Systems, 2016
	
	
	\bibitem{ref29}
	Zhang O, Srinivasan K,\emph{Mudra: User-friendly Fine-grained Gesture Recognition using WiFi Signals}.\hskip 1em plus
	0.5em minus 0.4em\relax The, International. 2016
	
	\bibitem{ref30}
	Song, Jie and Pece, Fabrizio and Fanello, Sean Ryan and Izadi, Shahram and Keskin, Cem and Hilliges, Otmar,\emph{In-air gestures around unmodified mobile devices}.\hskip 1em plus
	0.5em minus 0.4em\relax ACM Symposium on User Interface Software and Technology. 2014
	
	\bibitem{ref1}
	Chen, Feng Sheng and Fu, Chih Ming and Huang, Chung Lin, \emph{Hand gesture recognition using a real-time tracking method and hidden Markov models}.\hskip 1em plus
	0.5em minus 0.4em\relax Image and Vision Computing, 2003.
	
	\bibitem{ref2}
	Yao, Yi and Li, Chang Tsun, \emph{Hand gesture recognition and spotting in uncontrolled environments based on classifier weighting}.\hskip 1em plus
	0.5em minus 0.4em\relax IEEE International Conference on Image Processing, 2015.
	
	\bibitem{ref3}
	Ren, Zhou and Meng, Jingjing and Yuan, Junsong, \emph{Depth camera based hand gesture recognition and its applications in Human-Computer-Interaction}.\hskip 1em plus
	0.5em minus 0.4em\relax Communications and Signal Processing, 2011.
	
	\bibitem{ref4}
	Suarez, Jesus and Murphy, Robin R, \emph{Hand gesture recognition with depth images: A review}.\hskip 1em plus
	0.5em minus 0.4em\relax Ro-Man, 2012.
	
	\bibitem{ref5}
	Przybyla, Richard J and Tang, HaoYen and Guedes, Andr{\'e} and Shelton, Stefon E and Horsley, David A and Boser, Bernhard E, \emph{3D ultrasonic rangefinder on a chip}.\hskip 1em plus
	0.5em minus 0.4em\relax IEEE Journal of Solid-State Circuits, 2015.
	
	\bibitem{ref6}
	Li, Gang and Zhang, Rui and Ritchie, Matthew and Griffiths, Hugh, \emph{Sparsity-based dynamic hand gesture recognition using micro-Doppler signatures}.\hskip 1em plus
	0.5em minus 0.4em\relax Radar Conference, 2017.
	
	\bibitem{ref7}
	Kim, Youngwook and Toomajian, Brian, \emph{Hand Gesture Recognition Using Micro-Doppler Signatures With Convolutional Neural Network}.\hskip 1em plus
	0.5em minus 0.4em\relax IEEE Access, 2016.
	
	\bibitem{ref8}
	Molchanov, Pavlo and Gupta, Shalini and Kim, Kihwan and Pulli, Kari, \emph{Short-range FMCW monopulse radar for hand-gesture sensing}.\hskip 1em plus
	0.5em minus 0.4em\relax Radar Conference, 2015.
	
	\bibitem{ref9}
	Peng, Zhengyu and Li, Changzhi and Mu$\tilde{n}$oz-Ferreras, Jos$\acute{e}$ Mar$\acute{i}$a and G$\acute{o}$mez-Garc$\acute{i}$a, Roberto, \emph{An FMCW radar sensor for human gesture recognition in the presence of multiple targets}.\hskip 1em plus
	0.5em minus 0.4em\relax Microwave Bio Conference, 2017.
	
	\bibitem{ref10}
	Gupta, Sidhant and Morris, Daniel and Patel, Shwetak and Tan, Desney, \emph{SoundWave:using the doppler effect to sense gestures}.\hskip 1em plus
	0.5em minus 0.4em\relax 2012.
	
	\bibitem{ref12}
	Wang, Saiwen and Song, Jie and Lien, Jaime and Poupyrev, Ivan and Hilliges, Otmar, \emph{Interacting with Soli: Exploring Fine-Grained Dynamic Gesture Recognition in the Radio-Frequency Spectrum}.\hskip 1em plus
	0.5em minus 0.4em\relax Symposium on User Interface Software and Technology, 2016.
	
	\bibitem{ref13}
	Poupyrev, Ivan and Poupyrev, Ivan and Poupyrev, Ivan and Poupyrev, Ivan and Poupyrev, Ivan and Poupyrev, Ivan and Poupyrev, Ivan and Poupyrev, Ivan, \emph{Soli: ubiquitous gesture sensing with millimeter wave radar}.\hskip 1em plus
	0.5em minus 0.4em\relax Acm Transactions on Graphics, 2016.
	
	\bibitem{ref14}
	Graves, Alex and Mohamed, Abdel Rahman and Hinton, Geoffrey, \emph{Speech recognition with deep recurrent neural networks}.\hskip 1em plus
	0.5em minus 0.4em\relax IEEE International Conference on Acoustics, Speech and Signal Processing, 2013.
	
	\bibitem{ref15}
	Krizhevsky, Alex and Sutskever, Ilya and Hinton, Geoffrey E, \emph{ImageNet classification with deep convolutional neural networks}.\hskip 1em plus
	0.5em minus 0.4em\relax Communications of the ACM, 2017.
	
	\bibitem{ref16}
	Vu, Ngoc Thang and Adel, Heike and Gupta, Pankaj and Sch\"utze, Hinrich, \emph{Combining Recurrent and Convolutional Neural Networks for Relation Classification}.\hskip 1em plus
	0.5em minus 0.4em\relax NAACL, 2016.
	
	\bibitem{ref17}
	Kingma, Diederik P and Ba, Jimmy, \emph{Adam: A Method for Stochastic Optimization}.\hskip 1em plus
	0.5em minus 0.4em\relax Computer Science, 2014.
	
	\bibitem{ref18}
	Snapdragon, \emph{Snapdragon sense ID}.\hskip 1em plus
	0.5em minus 0.4em\relax {\url{https://www.qualcomm.com/products/features/security/fingerprint-sensors}}, 2015.
	
	\bibitem{ref19}
	HUG, \emph{HUG music player controlling}.\hskip 1em plus
	0.5em minus 0.4em\relax {\url{https://www.youtube.com/watch?v=8FgdiIb9WqY}}, 2016.
	
	\bibitem{ref20}
	Bosch, Anna and Zisserman, Andrew and Munoz, Xavier,\emph{Image Classification using Random Forests and Ferns}.\hskip 1em plus
	0.5em minus 0.4em\relax IEEE  International Conference on Computer Vision, 2007.
	
	\bibitem{ref31}
	óscar Pére, Piccardi M, García J, \emph{óscar Pére, Piccardi M, García J, et al. Comparison Between Genetic Algorithms and the Baum-Welch Algorithm in Learning HMMs for Human Activity Classification}.\hskip 1em plus
	0.5em minus 0.4em\relax Lecture Notes in Computer Science, 2007.
	
	\bibitem{ref32}
	MerrillIvan Skolnik, \emph{Introduction to radar systems}.\hskip 1em plus
	0.5em minus 0.4em\relax 1970.
\end{thebibliography}

\newpage

\begin{IEEEbiography}[{\includegraphics[width=1in,height=1.25in,clip,keepaspectratio]{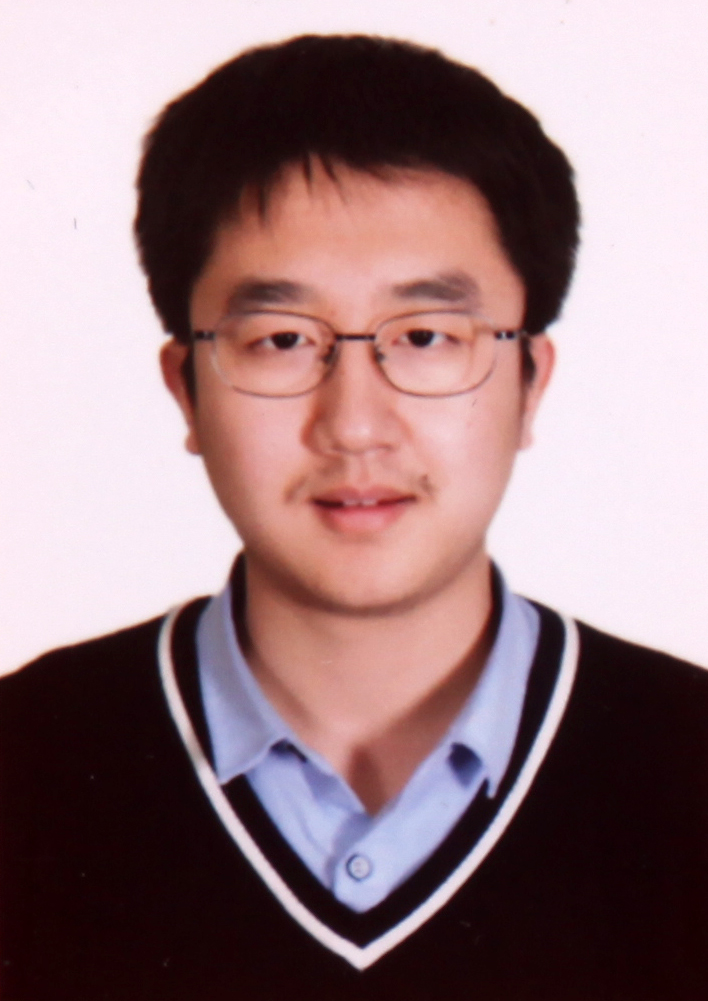}}]{Yu Sang}
	received the B.S. degree in electronic engineering from Tsinghua University, Beijing, China, in 2016.
	He is currently a master student in the Intelligence Sensing Lab. (ISL), Department of Electronic Engineering, Tsinghua University.
	His current research interests include signal processing and machine learning.
\end{IEEEbiography}

\begin{IEEEbiography}[{\includegraphics[width=1in,height=1.25in,clip,keepaspectratio]{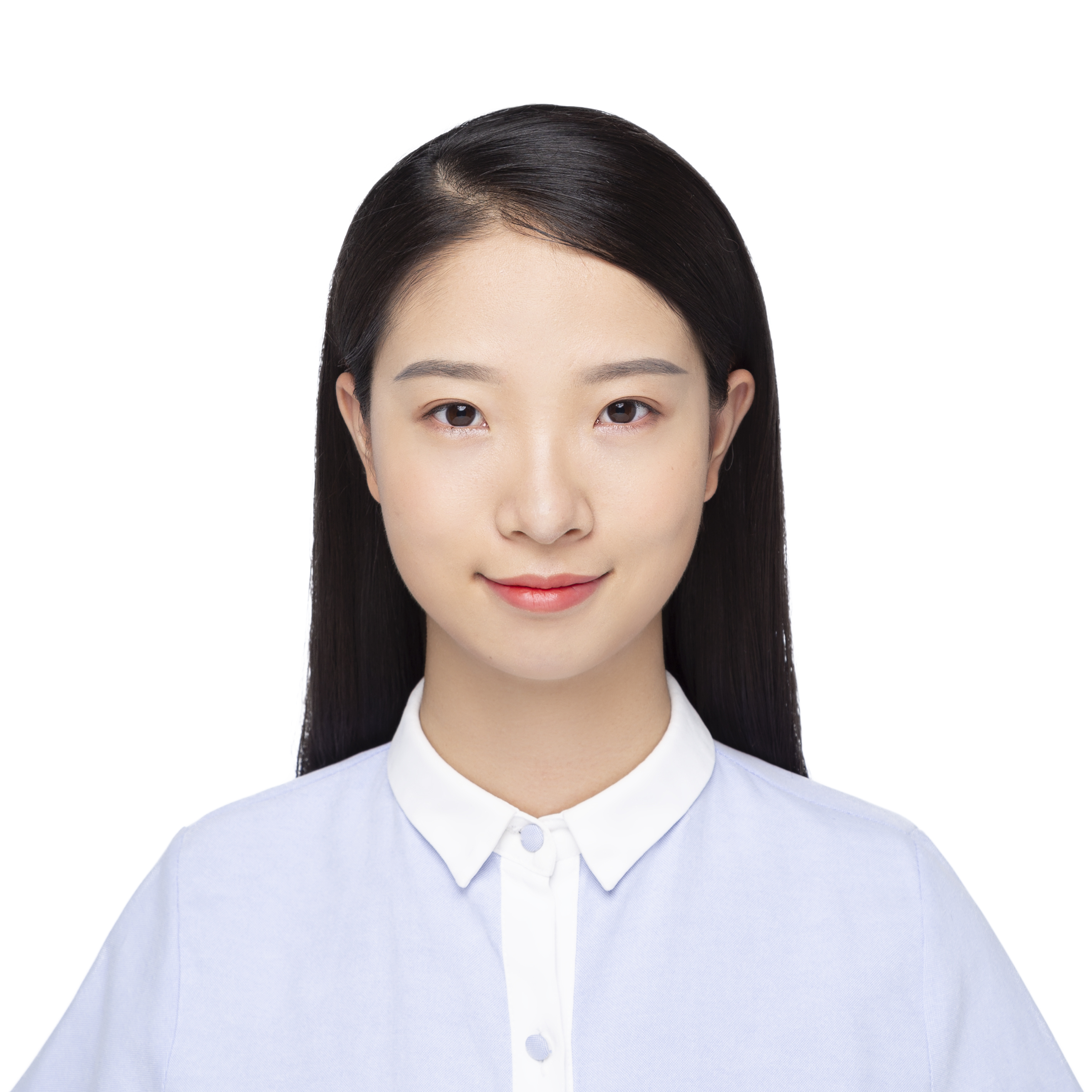}}]{Laixi Shi}
	will receive the B.S. degree in electronic engineering from Tsinghua University, Beijing, China, in 2018.
	She will be a Ph. D. student in the Department of Electrical and Computer Engineering, Carnegie Mellon University.
	Her current research interests include signal processing, nonconvex optimization and relative application.
\end{IEEEbiography}

\begin{IEEEbiography}
	[{\includegraphics[width=1in,height=1.25in,clip,keepaspectratio]{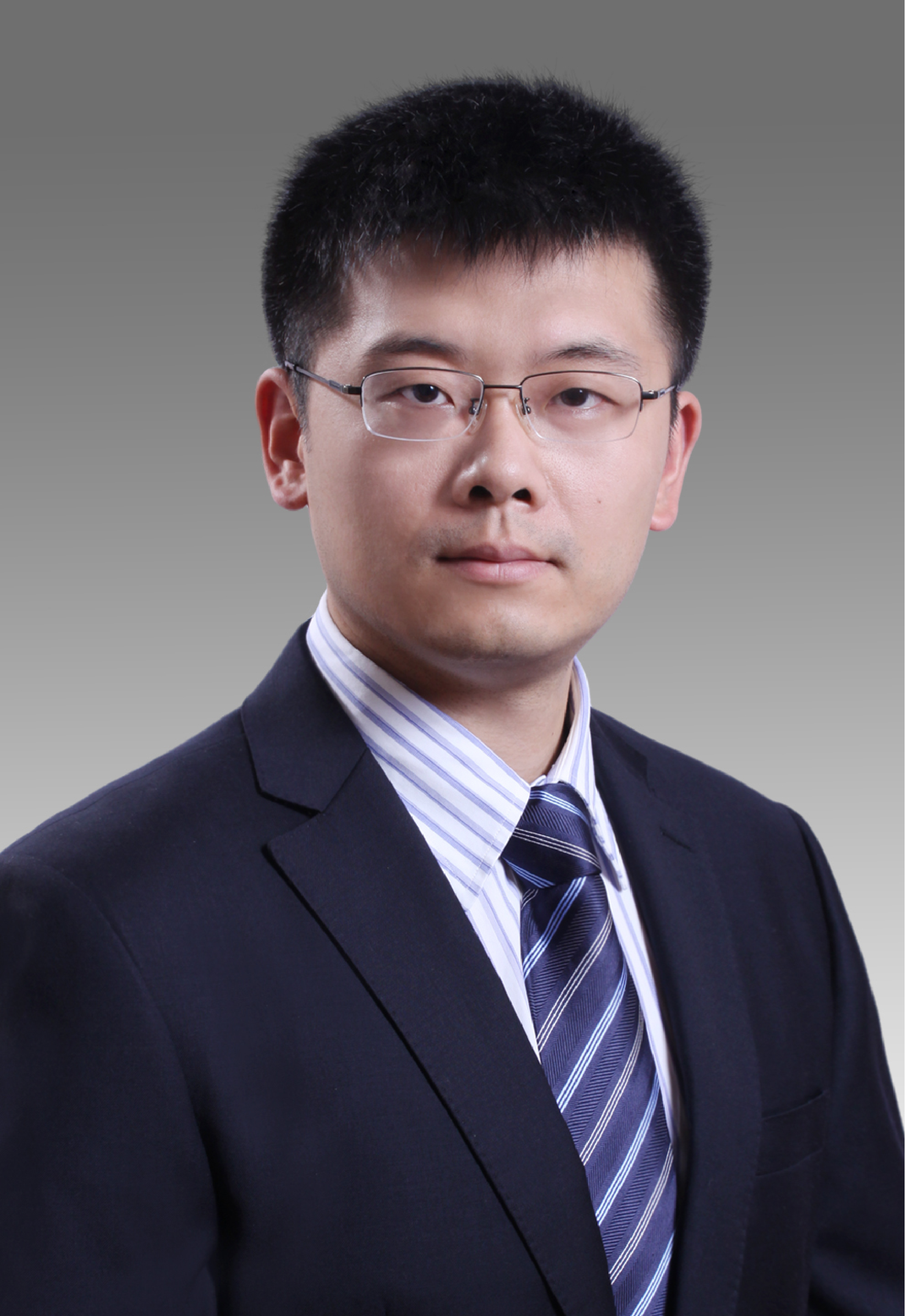}}]
	{Yimin Liu}
	(M'12) received the B.S. and Ph.D degrees (both with honors) in electronics engineering from the Tsinghua University, China, in 2004 and 2009, respectively.
	
	From 2004, he was with the Intelligence Sensing Lab. (ISL), Department of Electronic Engineering, Tsinghua University. He is currently an associate professor with Tsinghua, where his field of activity is study on new concept radar and other microwave sensing technologies. His current research interests include radar theory, statistic signal processing, compressive sensing and their applications in radar, spectrum sensing and intelligent transportation systems.
\end{IEEEbiography}

\EOD

\end{document}